\documentstyle[epsfig]{article}

\newcommand{\sfrac}[2]{\mbox{\footnotesize $\frac{#1}{#2}$}}

\begin{document}

\begin{center}
DYSON SCHWINGER EQUATIONS:\\ CONNECTING SMALL AND LARGE LENGTH-SCALES\\
\smallskip
CRAIG ROBERTS\\
{\em Physics Division, 203, Argonne National Laboratory, Argonne IL 60439, USA}
\end{center}

\hspace*{-\parindent} I illustrate the phenomenological application of
Dyson-Schwinger equations to the calculation of meson properties observable
at TJNAF.  Particular emphasis is given to this framework's ability to unify
long-range effects constrained by chiral symmetry with short-range effects
prescribed by perturbation theory, and interpolate between them.

\smallskip

\hspace*{-\parindent}{\it PACS numbers}: 12.38.Lg, 13.40.Gp, 14.40.Aq,
24.85.+p \vspace*{0.1em}

\hspace*{-\parindent}{\it Keywords}: Dyson-Schwinger Equations,
Nonperturbative methods, Continuum QCD, Confinement, Dynamical chiral
symmetry breaking, Pion electromagnetic and anomalous transition form factors
\medskip

\begin{center}
{\large\it 1.~Dressed Quarks}
\end{center}

The Dyson-Schwinger equations (DSEs)~[\ref{cdragw}] provide a nonperturbative
approach to studying the continuum formulation of QCD, making accessible
phenomena such as confinement, dynamical chiral symmetry breaking (DCSB) and
bound state structure.  However, they also provide a generating tool for
perturbation theory and hence their phenomenological application is tightly
constrained at high-energy.  This is the particular feature of the
phenomenological application of DSEs: their ability to furnish a unified
description of high- and low-energy phenomena in QCD.  It is elucidated in
Refs.~\ref{peter},\ref{cdranu} and here I only illustrate this aspect, using
as primary exemplars the electromagnetic pion form factor and the
$\gamma^\ast \pi \to \gamma$ transition form factor, which are particularly
relevant to the TJNAF community.

A key element in the description of hadronic observables is the dressing of
the quark propagator, which is described by the quark DSE:
\begin{eqnarray}
\label{genS}
S_f(p)^{-1} & := & 
i \gamma\cdot p \,A_f(p^2) + B_f(p^2)  =
A_f(p^2) \left( i \gamma\cdot p + M_f(p^2) \right)\\
\label{gendse} & = & Z_2 (i\gamma\cdot p + m_f^{\rm bm})
+\, Z_1\, \int^\Lambda_q \,
g^2 D_{\mu\nu}(p-q) \frac{\lambda^a}{2}\gamma_\mu S_f(q)
\Gamma^{fa}_\nu(q,p) .
\end{eqnarray}
Here $f$ is a flavour label, $D_{\mu\nu}(k)$ is the dressed-gluon
propagator~[\ref{pennington},\ref{alkofer}], $\Gamma^{fa}_\nu(q,p)$ is the
dressed-quark-gluon vertex, $m_f^{\rm bm}$ is the $\Lambda$-dependent
current-quark bare mass and $\int^\Lambda_q := \int^\Lambda d^4 q/(2\pi)^4$
represents mnemonically a {\em translationally-invariant} regularisation of
the integral, with $\Lambda$ the regularisation mass-scale.  The
quark-gluon-vertex and quark wave function renormalisation constants, $Z_1$
and $Z_2$, depend on the renormalisation point, $\zeta$, and the
regularisation mass-scale.

The qualitative features of the solution of Eq.~(\ref{gendse}) are known.  In
QCD the chiral limit is defined by $\hat m = 0$, where $\hat m$ is the
renormalisation-point-independent current-quark mass.  For $\hat m = 0$ there
is no mass-like divergence in the perturbative evaluation of the quark self
energy and hence for $p^2>20\,$GeV$^2$ the solution of Eq.~(\ref{gendse})
is~[\ref{mr97}]
\begin{equation}
\label{Mchiral}
M_0(p^2) \stackrel{{\rm large}-p^2}{=}\,
\frac{2\pi^2\gamma_m}{3}\,\frac{\left(-\,\langle \bar q q \rangle^0\right)}
           {p^2
        \left(\sfrac{1}{2}\ln\left[p^2/\Lambda_{\rm QCD}^2\right]
        \right)^{1-\gamma_m}}\,,
\end{equation}
where $\gamma_m=12/(33-2 N_f)$ is the gauge-independent mass anomalous
dimension and $\langle \bar q q \rangle^0$ is the
renormalisation-point-independent vacuum quark condensate.  The
momentum-dependence is a model-independent result.  The existence of DCSB
means that $\langle \bar q q \rangle^0 \neq 0$, however, its actual value
depends on the long-range behaviour of $D_{\mu\nu}(k)$ and
$\Gamma^{0a}_\nu(q,p)$, which is modelled in contemporary DSE studies.
Requiring a good description of light-meson observables necessitates $\langle
\bar q q \rangle^0\approx - (0.24\,\mbox{GeV})^3$.

In contrast, for $\hat m_f \neq 0$,
\begin{equation}
\label{masanom}
M_f(p^2) \stackrel{{\rm large}-p^2}{=} \frac{\hat m_f}
{\left(\sfrac{1}{2}\ln\left[p^2/\Lambda_{\rm
QCD}^2\right]\right)^{\gamma_m}}\,.
\end{equation}
An obvious qualitative difference is that, relative to Eq.~(\ref{masanom}),
the chiral-limit solution is $1/p^2$-suppressed in the ultraviolet.

\begin{figure}[t]
\centering{\
\epsfig{figure=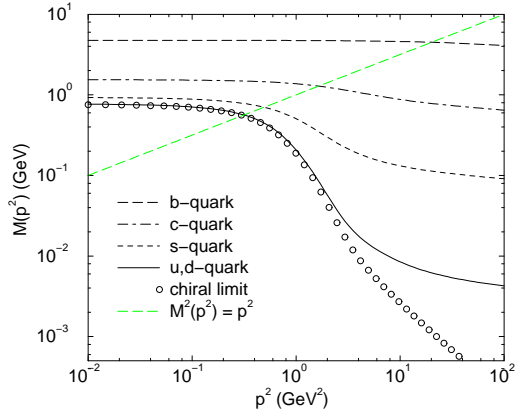,height=5.5cm}}
\caption{Quark mass function obtained as a solution of
Eq.~(\protect\ref{gendse}) using $D_{\mu\nu}(k)$ and $\Gamma^{fa}_\nu(q,p)$
from Ref.~\protect\ref{mr97}, and $m_{u,d}^\zeta = 3.7\,$MeV, $m_s^\zeta =
82\,$MeV, $m_c^\zeta=0.58\,$GeV and $m_b^\zeta=3.8\,$GeV ($\zeta= 19\,$GeV).
The indicated solutions of $M^2(p^2)=p^2$ define the Euclidean
constituent-quark mass, $M^E_f$, which takes the values:
$M^E_{u,d}=0.56\,$GeV, $M^E_s=0.70\,$GeV, $M^E_c= 1.3\,$GeV, $M^E_b=
4.6\,$GeV.
\label{mp2fig}}
\end{figure}
There is some quantitative model-dependence in the $p^2$-evolution of
$M_f(p^2)$ into the infrared.  However, for any forms of $D_{\mu\nu}$ and
$\Gamma^{fa}_\nu$ that provide an accurate description of $f_{\pi,K}$ and
$m_{\pi,K}$, one obtains profiles like those illustrated in
Fig.~\ref{mp2fig}.  The evolution to coincidence between the chiral-limit and
$u,d$-quark mass functions, apparent in this figure, makes clear the
transition from the perturbative to the nonperturbative domain.  The chiral
limit mass-function is nonzero {\it only} because of the nonperturbative DCSB
mechanism whereas the $u,d$-quark mass function is purely perturbative at
$p^2>20\,$GeV$^2$, where Eq.~(\ref{masanom}) is accurate.  The DCSB mechanism
thus has a significant effect on the propagation characteristics of
$u,d,s$-quarks, and this is fundamentally important in QCD with observable
consequences.

\medskip

\begin{center}
{\large\it 2.~Bound States}
\end{center}

A meson is a bound state of a dressed-quark and -antiquark, and its internal
structure is described by a Bethe-Salpeter amplitude obtained as the solution
of
\begin{eqnarray}
\label{genbse}
\left[\Gamma_P(k;Q)\right]_{tu} &= & 
\int^\Lambda_q \,[\chi_P(q;Q)]_{sr} \,K^{rs}_{tu}(q,k;Q)\,,
\end{eqnarray}
where $\chi_P(q;Q) = {\cal S}(q_+) \Gamma_P(q;Q) {\cal S}(q_-)$; ${\cal S}(q)
= {\rm diag}(S_u(q),S_d(q),S_s(q), \ldots)$; $q_+=q + \eta_Q\, Q$, $q_-=q -
(1-\eta_Q)\, Q$, with $Q$ the total momentum of the bound state and
$\eta_Q\in[0,1]$ the relative-momentum partitioning parameter; and
$r$,\ldots,$u$ represent colour-, Dirac- and flavour-matrix indices.  For a
pseudoscalar meson, such as the pion, the solution has the general form
\begin{eqnarray}
\label{genpibsa}
\Gamma_P(k;Q) & = &  T^P \gamma_5 \left[ i E_P(k;Q) + 
\gamma\cdot Q F_P(k;Q) \rule{0mm}{5mm}\right. \\
\nonumber & & 
\left. \rule{0mm}{5mm}+ \gamma\cdot k \,k \cdot Q\, G_P(k;Q) 
+ \sigma_{\mu\nu}\,k_\mu Q_\nu \,H_P(k;Q) 
\right]\,,
\end{eqnarray}
where $T^P$ is a flavour matrix identifying the meson; e.g., $T^{\pi^+}=
\sfrac{1}{2}\left(\lambda^1 + i \lambda^2\right)$, with
$\{\lambda^j,j=1\ldots N_f^2-1\}$ the Gell-Mann matrices of $SU(N_f)$.

In Eq.~(\ref{genbse}), $K$ is the renormalised, fully-amputated,
quark-antiquark scattering kernel.  Important in the successful application
of DSEs is that $K$ has a systematic skeleton expansion in terms of the
elementary dressed-particle Schwinger functions; e.g., the dressed-quark and
-gluon propagators.  The expansion introduced in Ref.~\ref{bender96} provides
a means of constructing a kernel that, order-by-order in the number of
vertices, ensures the preservation of vector and axial-vector Ward-Takahashi
identities; i.e., current conservation.  Only with such a truncation is an
accurate description of the light-quark mesons possible.

In QCD the leptonic decay constant of a pseudoscalar meson is~[\ref{mr97}]
\begin{eqnarray}
\label{fwk}
f_{P} \,Q_\mu 
= {\rm tr}\,Z_2 \int^\Lambda_k \,(T^P)^{\rm T}\gamma_5\gamma_\mu\,
        \chi_P(k;Q),
\end{eqnarray}
where the trace is over colour, Dirac and flavour indices.
Equation~(\ref{fwk}) is {\it exact}: the $\Lambda$-dependence of $Z_2$
ensures that the right-hand-side (r.h.s.) is finite as $\Lambda \to \infty$,
and its $\zeta$- and gauge-dependence is just that necessary to compensate
that of $\chi_P(k;Q)$.

In the chiral limit the axial-vector current is conserved, and employing any
Ward-Takahashi identity preserving truncation of $K$ one obtains
\begin{equation}
\label{bwti}
\begin{array}{ll}
f_P E_P(k;0)  =   B_0(k^2)\,,\; & 
F_R(k;0) +  2 \, f_P F_P(k;0)   =  A_0(k^2)\,, \\
G_R(k;0) +  2 \,f_P G_P(k;0)     =  2 A_0^\prime(k^2)\,,\; &
H_R(k;0) +  2 \,f_P H_P(k;0)     =  0\,,
\end{array}
\end{equation}
where $F_R$, $G_R$ and $H_R$ are calculable functions in the dressed
axial-vector vertex, $\Gamma_{5 \mu}^H(k;Q)$.  These identities are
associated with Goldstone's theorem and in fact one can show~[\ref{mr97}]
that when chiral symmetry is dynamically broken: 1) the flavour-nonsinglet,
pseudoscalar BSE has a massless solution; 2) the Bethe-Salpeter amplitude for
the massless bound state has a term proportional to $\gamma_5$ alone, with
the momentum-dependence of $E_P(k;0)$ completely determined by that of
$B_0(k^2)$, in addition to terms proportional to other pseudoscalar Dirac
structures that are nonzero; and 3) $\Gamma_{5 \mu}^P(k;Q)$ is dominated by
the pseudoscalar bound state pole for $Q^2\simeq 0$.  The converse is also
true.  Hence, in the chiral limit, the pion is a massless composite of a
quark and an antiquark, each of which has an effective mass $M^E \sim
0.5\,$GeV.

For nonzero values of the current-quark mass, whether small or large, instead
of Eqs.~(\ref{bwti}) one obtains~[\ref{mr97}]
\begin{equation}
\label{gmora}
f_P^2\,m_P^2  =  \,- \,{\cal M}_P\,\langle \bar q q \rangle^P_\zeta ,
\end{equation}
where
${\cal M}_P = {\rm tr}_{\rm f}
\left[M_{(\zeta)}\,\left\{T^P,\left(T^P\right)^{\rm T}\right\}\right]$
e.g., for the $\pi$: ${\cal M}_{\pi^+} = m_u^\zeta + m_d^\zeta\,$, and
\begin{eqnarray}
\label{inhadqbq}
-\langle \bar q q \rangle^P_\zeta & = & 
f_P \,{\rm tr} \,
Z_4\int_q^\Lambda\, \left(T^P\right)^{\rm T} \gamma_5 \chi_P(q;Q) .
\end{eqnarray}
Equation~(\ref{gmora}) is an exact mass formula for flavour non-singlet
mesons and I note that the r.h.s.  does not involve a difference of massive
quark propagators: a phenomenological assumption often employed.  $\langle
\bar q q \rangle^P_\zeta$ in Eq.~(\ref{inhadqbq}) is an ``in-hadron''
condensate.  It is gauge-independent and its renormalisation point dependence
is exactly that required to ensure that the r.h.s. of Eq.~(\ref{gmora}) is
renormalisation point {\it independent}.

For small current-quark masses, $\hat m_q \sim 0$, Eq.~(\ref{gmora}) yields
what is commonly called the Gell-Mann--Oakes--Renner relation; i.e., $m_P^2
\propto \hat m_q$, because
\begin{equation}
\lim_{\hat m \to 0} \langle \bar q q \rangle^P_\zeta = 
\langle \bar q q \rangle^0_\zeta\,.
\end{equation}
However, it also has an important corollary when the current-mass, $\hat
m_Q$, of one or both constituents becomes large, predicting~[\ref{misha}]
\begin{equation}
m_P \propto \hat m_Q\,,
\end{equation}
which follows because $\langle \bar q q \rangle^P_\zeta$ is $\hat
m_Q$-independent for large-$\hat m_Q$ and $f_P \propto m_P^{-1/2}$.  The
transition from the quadratic to the linear mass-relation occurs at $\hat m_q
\approx 2\,\hat m_s$~[\ref{pieterrostock}], at which point explicit chiral
symmetry breaking overwhelms DCSB.

Two other important model-independent results can be obtained~[\ref{mr97}]
from Eq.~(\ref{genbse}) and the systematic construction of $K$.  The scalar
functions in Eq.~(\ref{genpibsa}) depend on three invariants; e.g.,
$E_P(k;Q)= E_P(k^2,k\cdot Q,Q^2)$.  The zeroth Chebyshev moment of these
functions; e.g.,
\begin{equation}
^0\! E_P(k^2,Q^2):= \frac{2}{\pi}\,\int_0^\pi\,dx\,\sqrt{1-x^2}\, E_P(k;Q)\,,
        \; k\cdot Q := x \sqrt{k^2 Q^2}
\end{equation}
are dominant in the description of bound state properties, and 
\begin{equation}
^0\! E_P(k^2,Q^2) \stackrel{{\rm large}-k^2}{\propto}\, M_0(k^2)\,,
\end{equation}
with $^0\! F_P(k^2,Q^2)$, $k^2\, ^0\! G_P(k^2,Q^2)$ and $k^2\, ^0\!
H_P(k^2,Q^2)$ behaving in precisely the same way.  Further
\begin{equation}
\label{uvrel}
k^2\,^0\!G_P(k^2,Q^2) \stackrel{{\rm large}-k^2}{=}\,
2 \,^0\! F_P(k^2,Q^2)\,.
\end{equation}
These results determine the asymptotic form of the electromagnetic pion form
factor.

\medskip

\begin{figure}[t]
\centering{\ \epsfig{figure=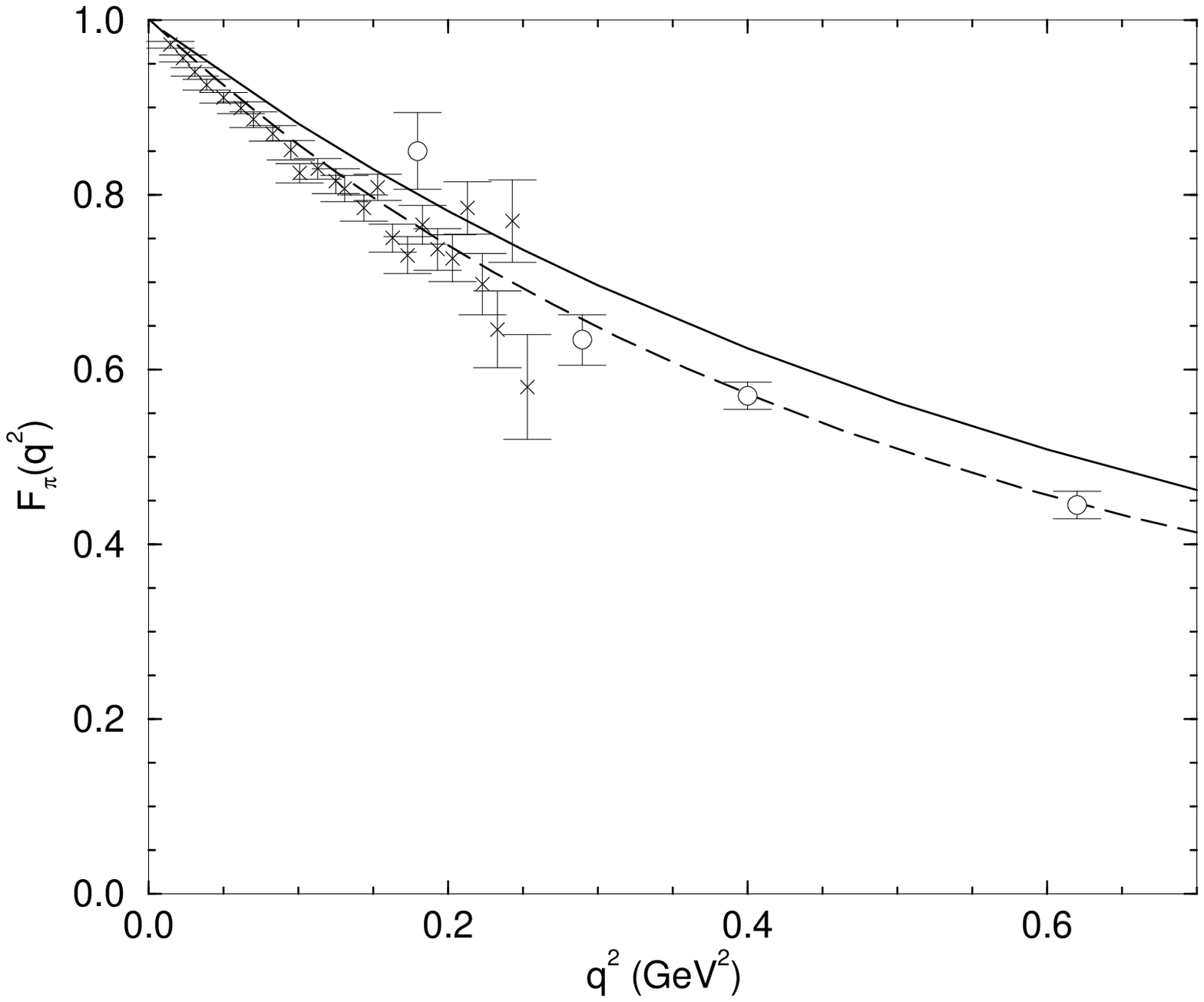,height=5.2cm}}\vspace*{0.5\baselineskip}

\centering{\ \epsfig{figure=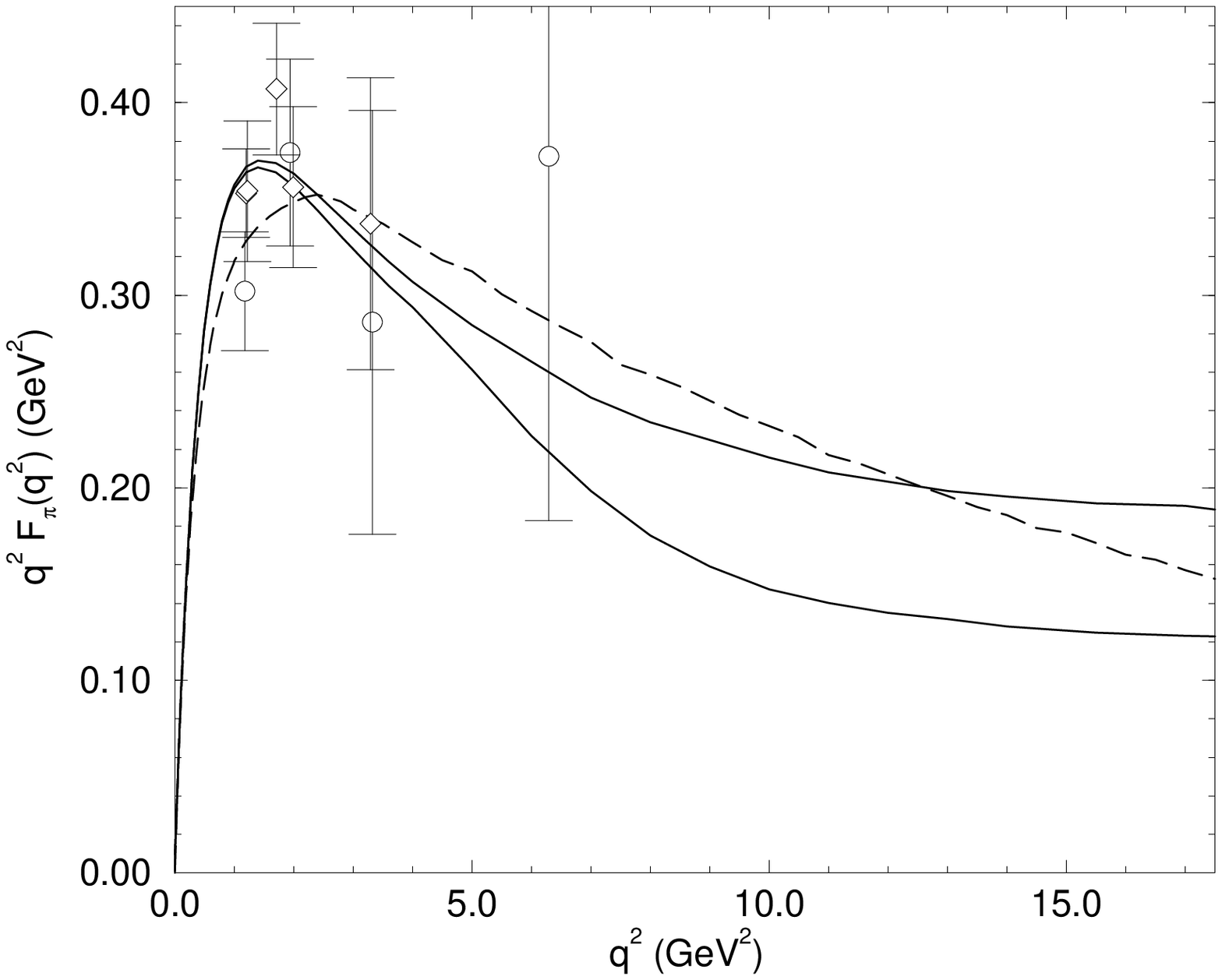,height=5.2cm}}
\caption{Upper panel: calculated pion form factor compared with data at
small-$q^2$.  Lower panel: the large-$q^2$ comparison, with the two solid
lines showing the range of model-dependent uncertainty.  In both panels the
dashed line~[\protect\ref{cdrpion}] assumes that $F_\pi = 0 = G_\pi = H_\pi$,
and the data are taken from Refs.~\protect\ref{piffdat}.
\label{piff}}
\end{figure}
\begin{center}
{\large\it 3.~Electromagnetic Pion Form Factor}
\end{center}

The impulse approximation to the electromagnetic pion form factor
is~[\ref{pmpion}]
\begin{eqnarray}
\label{pipiA}
\lefteqn{(p_1 + p_2)_\mu\,F_\pi(q^2):= \Lambda_\mu(p_1,p_2)  }\\
& & \nonumber
= 2 N_c\,{\rm tr}_D \int_k^\Lambda \bar\Gamma_\pi(k;-p_2)
S(k_{++})\, i\Gamma^\gamma_\mu(k_{++},k_{+-})\,S(k_{+-})\,
\Gamma_\pi(k_{0-};p_1)\,S(k_{--}),
\end{eqnarray}
where $\bar\Gamma_\pi(q;-P)^{\rm T} := C^{-1}\,\Gamma_\pi(-q;-P)\,C$ with
$C=\gamma_2\gamma_4$, the charge conjugation matrix, and $k_{\alpha\beta}:= k
+ \alpha p_1/2 + \beta q/2$, $p_2:= p_1 + q$.  No renormalisation constants
appear explicitly in Eq.~(\ref{pipiA}) because the renormalised
dressed-quark-photon vertex, $\Gamma^\gamma_\mu$, satisfies the vector
Ward-Takahashi identity.  This also ensures current conservation:
$(p_1-p_2)_\mu\,\Lambda_\mu(p_1,p_2)=0$.
The calculation of $F_\pi(q^2)$ is simplified by using a model algebraic
parametrisation of the dressed-quark propagator that efficiently
characterises the essential elements of the solution of the quark-DSE and
determines the pion Bethe-Salpeter amplitude via Eqs.~(\ref{bwti}) and
(\ref{uvrel}), and an efficacious {\it Ansatz}~[\ref{bc80}] for
$\Gamma^\gamma_\mu$ in which the vertex is completely determined by the
dressed-quark propagator.  The result is depicted in Fig.~\ref{piff}.  The
current uncertainty in the experimental data at intermediate $q^2$ is
apparent in the lower panel, as is the difference between the results
calculated with or without the pseudovector components: $F$, $G$, of the pion
Bethe-Salpeter amplitude.  These components provide the dominant contribution
to $F_\pi(q^2)$ at large pion energy~[\ref{pmpion}] because of the
multiplicative factors: $\gamma\cdot Q$ and $\gamma\cdot k\,k\cdot Q$, which
contribute an additional power of $q^2$ in the numerator of those terms
involving $F^2$, $FG$ and $G^2$ relative to those proportional to $E$.
Including them one finds
\begin{equation}
\label{FUV}
F_\pi(q^2) \stackrel{{\rm large}-q^2}{\propto} \frac{\alpha(q^2)}{q^2}\,
        \frac{(-\langle \bar q q\rangle^0_{q^2})^2}{f_\pi^4}\,;
\end{equation}
i.e., $q^2 F_\pi(q^2) \approx {\rm const.}$, up to calculable $\ln
q^2$-corrections, in agreement with the expectations raised by perturbative
QCD.  If the pseudovector components of $\Gamma_\pi$ are neglected, the
additional numerator factor of $q^2$ is missing and one
obtains~[\ref{cdrpion}] $q^4 F_\pi(q^2)\approx {\rm const.}$.

\medskip

\begin{center}
{\large\it 4.~$\gamma^\ast \pi \to \gamma$ Transition Form Factor}
\end{center}

The impulse approximation to this form factor is 
\begin{eqnarray}
\label{anomaly}
\lefteqn{T^3_{\mu\nu}(k_1,k_2) := 
\frac{1}{4\pi^2} i\varepsilon_{\mu\nu\rho\sigma}\,k_{1\rho} k_{2\rho} 
        \hat T(k_1^2,k_1\cdot k_2,k_2^2) }\\
& & \nonumber 
 = {\rm tr}\int^\Lambda_q 
S(q_1)\,        \Gamma_\pi(\hat q;-P) 
S(q_2)\,i{\cal Q}\Gamma^\gamma_\mu(q_2,q_{12})
                S(q_{12})\,i{\cal Q}\Gamma^\gamma_\nu(q_{12},q_1) \,,
\end{eqnarray}
where ${\cal Q}=(\sfrac{1}{3} I + \tau^3)/2 = {\rm diag}(2/3,-1/3)$, and
$k_1$, $k_2$ are the photon momenta [on-shell: $k_1^2=0=k_2^2$, $2 k_1\cdot
k_2=P^2$], $q$ is the loop-momentum, and $q_1:= q-k_1$, $q_2:= q+k_2$, $\hat
q:= \sfrac{1}{2}(q_1+q_2)$, $q_{12}:= q-k_1+k_2$.  The manner in which the
``triangle anomaly'' is recovered with no dependence on model parameters is
described in Ref.~[\ref{pmpion}], and this same mechanism applies to all
anomalous pion and photopion processes~[\ref{anomalies}].  It is a unique
feature of the DSE framework.

Using this expression one can calculate~[\ref{mitchell},\ref{klabucar}] $\hat
T(k_1^2,k_1\cdot k_2,k_2^2)$ when one or both of the photons is off shell and
also determine the asymptotic behaviour analytically.  The formal character
of that derivation is presented in Ref.~\ref{klabucar} but, in neglecting
essential aspects of renormalisation, it is imprecise.  I remedy that here
for the illustrative case of one photon off-shell: $k_1^2= Q^2$.

At large-$Q^2$, $k_1\cdot k_2 \approx -Q^2/2$, and in Eq.~(\ref{anomaly}) the
pion Bethe-Salpeter amplitude focuses the integration support at $\hat q=0$.
As a consequence the asymptotic behaviour of the integral can be determined
using 
\begin{equation}
S(q_{12}) \approx  \frac{1}{Z_2} \frac{1}{Q^2}i\gamma\cdot (k_1-k_2) \,,\;
        \Gamma^\gamma_\mu \approx Z_1 \,\gamma_\mu\,,
\end{equation}
which follow from Eq.~(\ref{gendse}) and the DSE for the quark-photon vertex,
so that
\begin{equation}
T^3_{\mu\nu}(k_1,k_2) \approx i\frac{1}{Q^2}\,(k_1-k_2)_\sigma
\,{\rm tr}\,Z_2\int^\Lambda_q \sfrac{1}{6}\tau^3\,
\chi_\pi(\hat q;-P) \gamma_\mu \gamma_\sigma \gamma_\nu\,,
\end{equation}
where I have used the Ward identity: $Z_1 = Z_2$.  Hence, the transition form
factor
\begin{eqnarray}
\nonumber
\lefteqn{
\frac{1}{4\pi^2} i\varepsilon_{\mu\nu\rho\sigma}\,k_{1\rho} k_{2\sigma}\, 
        T(Q^2) := 
T^3_{\mu\nu}(k_1,k_2) +T^3_{\nu\mu}(k_2,k_1)  }\\
& \approx &
-\frac{1}{Q^2}\,
i\varepsilon_{\mu\nu\rho\sigma}\,(k_1-k_2)_\rho\,
\sfrac{2}{3}\,{\rm tr}\,Z_2\int^\Lambda_q \sfrac{1}{2}\tau^3\,
\chi_\pi(\hat q;-P) \gamma_5 \gamma_\sigma \\
& = & i\varepsilon_{\mu\nu\rho\sigma}\,k_{1\rho} k_{2\sigma}\,
        \frac{4}{3}\,\frac{f_\pi}{Q^2}\,,
\end{eqnarray}
where the last line follows from Eq.~(\ref{fwk}); i.e.,
\begin{equation}
T(Q^2) \stackrel{{\rm large}-Q^2}{\approx}
 \frac{4}{3}\,\frac{4\pi^2 f_\pi}{Q^2}\,,
\end{equation}
in agreement with the expectations raised by perturbative QCD.  Thus, as with
Eq.~(\ref{pipiA}), one equation unifies the small- and large-$Q^2$ results
and predicts the evolution between them~[\ref{mitchell}-\ref{peterD}].

\medskip

\begin{center}
{\large\it 5.~Epilogue}
\end{center}

I have been necessarily brief.  There are many other applications of interest
to this community, among them the diffractive electroproduction of neutral
vector mesons~[\ref{pichowsky}], the electromagnetic form factors of their
charged states~[\ref{hawes}] and the unification of light- and heavy-meson
observables~[\ref{misha}].  The most pressing contemporary challenge relevant
to this community is the extension of the framework to the calculation of
baryon observables, which is underway.

\smallskip

\begin{center}
Acknowledgments
\end{center}
I would like to thank Dubravko Klabucar and the organisers for their
assistance, kindness and hospitality.  This work was supported by the US
Department of Energy, Nuclear Physics Division, under contract number
W-31-109-ENG-38, and the National Science Foundation, under grant
no.~INT-9603385.

\begin{center}
References
\end{center}
\begin{enumerate}
\item \label{cdragw} C.D.~Roberts and A.G.~Williams,
Prog. Part. Nucl. Phys. {\bf 33} (1994) 477. \vspace*{-0.5\baselineskip}
%
\item \label{peter} P.C.~Tandy, Prog. Part. Nucl. Phys. {\bf 39} (1997)
117.\vspace*{-0.5\baselineskip}
%
\item \label{cdranu} C.D.~Roberts,
``Nonperturbative QCD with modern tools,"
nucl-th/9807026.\vspace*{-0.5\baselineskip} 
%
\item \label{pennington} M.R.~Pennington, ``Calculating hadronic properties
in strong QCD," hep-ph/9611242.\vspace*{-0.5\baselineskip} 
%
\item \label{alkofer} R.~Alkofer, S.~Ahlig and L.v.~Smekal, ``The infrared
behavior of gluon, ghost, and quark propagators in Landau gauge QCD,"
hep-ph/9901322, these proceedings.\vspace*{-0.5\baselineskip}
%
\item \label{mr97} P.~Maris and C.D.~Roberts, Phys. Rev. {\bf C56} (1997)
3369.\vspace*{-0.5\baselineskip} 
%
\item \label{bender96} A.~Bender, C.D.~Roberts and L. v. Smekal,
Phys. Lett. {\bf B380} (1996) 7.\vspace*{-0.5\baselineskip} 
%
\item \label{misha} M.A.~Ivanov, Y.L.~Kalinovsky and C.D.~Roberts, ``Survey
of heavy-meson observables", nucl-th/9812063.\vspace*{-0.5\baselineskip}
%
\item \label{pieterrostock} P.~Maris and C.D.~Roberts, ``Differences between
heavy and light quarks," nucl-th/9710062.\vspace*{-0.5\baselineskip}
%
\item \label{pmpion} P.~Maris and C.D.~Roberts, Phys. Rev. {\bf C58} (1998)
3659. \vspace*{-0.5\baselineskip}
%
\item \label{bc80} J.S.~Ball and T.~Chiu, Phys. Rev. {\bf D22} (1980)
2542.\vspace*{-0.5\baselineskip} 
%
\item \label{piffdat} S.R.~Amendolia {\it et al.}, Nucl. Phys. {\bf B277}
(1986) 168; 
C.J.~Bebek {\it et al.}, Phys. Rev. {\bf D13} (1976) 25;
C.J.~Bebek {\it et al.}, Phys. Rev. {\bf D17} (1978)
1693.\vspace*{-0.5\baselineskip} 
%
\item \label{cdrpion} C.D.~Roberts, Nucl. Phys. {\bf A605} (1996)
475.\vspace*{-0.5\baselineskip} 
%
\item \label{anomalies} C.D.~Roberts, R.T.~Cahill and J.~Praschifka,
Ann. Phys. {\bf 188} (1988) 20; 
R.~Alkofer and C.D.~Roberts, Phys. Lett. {\bf B 369} (1996)
101.\vspace*{-0.5\baselineskip} 
%
\item \label{mitchell} M.R.~Frank, K.L.~Mitchell, C.D.~Roberts and
P.C.~Tandy, Phys. Lett. {\bf B 359} (1995) 17.\vspace*{-0.5\baselineskip} 
%
\item \label{klabucar} D.~Kekez and D.~Klabucar, ``$\gamma^\ast$ $\gamma$
$\to$ $\pi^0$ transition and asymptotics of $\gamma^\ast \gamma$ and
$\gamma^\ast \gamma^\ast$ transitions of other unflavored pseudoscalar
mesons," hep-ph/9812495; and these proceedings.\vspace*{-0.5\baselineskip} 
%
\item \label{peterD} P.C. Tandy, these proceedings.\vspace*{-0.5\baselineskip} 
\item \label{pichowsky} M.A.~Pichowsky and T.S.~Lee, Phys. Rev. {\bf D56}
(1997) 1644.\vspace*{-0.5\baselineskip} 
%
\item \label{hawes} F.T.~Hawes and M.A.~Pichowsky, ``Electromagnetic
form-factors of light vector mesons,"
nucl-th/9806025.\vspace*{-0.5\baselineskip}
\end{enumerate}

\end{document}